\documentclass[sigconf]{acmart}
\AtBeginDocument{%
  }

\setcopyright{acmlicensed}
\copyrightyear{2026}
\acmYear{2026}
\acmDOI{XXXXXXX.XXXXXXX}
\acmConference[Conference acronym 'XX]{Make sure to enter the correct
  conference title from your rights confirmation email}{June 03--05,
  2018}{Woodstock, NY}
\acmISBN{978-1-4503-XXXX-X/2018/06}

\acmSubmissionID{123-A56-BU3}

\copyrightyear{2026}
\acmYear{2026}
\setcopyright{cc}
\setcctype{by}
\acmConference[CI '26]{Proceedings of the ACM Collective Intelligence Conference}{September 27--30, 2026}{Alexandria, VA, USA}
\acmBooktitle{Proceedings of the ACM Collective Intelligence Conference (CI '26), September 27--30, 2026, Alexandria, VA, USA}
\acmDOI{10.1145/3834581.3838628}
\acmISBN{979-8-4007-2895-2/2026/09}
\begin{document}

\title{Communication Heterogeneity and Collective Consensus in Neural Cellular Automata}

\author{Nishit Singh}
\email{f20221317@pilani.bits-pilani.ac.in}
\affiliation{%
  \institution{Birla Institute of Technology and Science, Pilani}
  \city{Pilani}
  \state{Rajasthan}
  \country{India}
}

\begin{abstract}
Reaching global agreement from purely local interactions is a defining problem of
collective intelligence, and most models of it assume that all agents share a
single communication protocol. We ask what happens when they do not. Using a
Neural Cellular Automaton in which a population of cells must solve the density
classification task, agreeing on a global majority that no individual can observe,
we introduce ``languages'' as sub-populations that read one another's messages
through a translation with a tunable ``linguistic distance''. We find that
linguistic distance slows consensus, that it produces mild divergence between
groups rather than full fragmentation, and that a collective whose shared rule was
trained under diverse protocols is robust to mismatch; a homogeneously trained
one is not. The findings hold on both a ring and a two-dimensional grid, and admit
a natural reading as Ising relaxation, in which a foreign-language region acts as a
boundary defect that leaves the system in a higher-energy, partially ordered state.
These patterns are qualitatively consistent with effects reported in human group studies, suggesting that distance between communication protocols is possibly a minimal mechanism sufficient to produce them, without anything language-specific.
\end{abstract}


\begin{teaserfigure}
 \includegraphics[width=\textwidth]{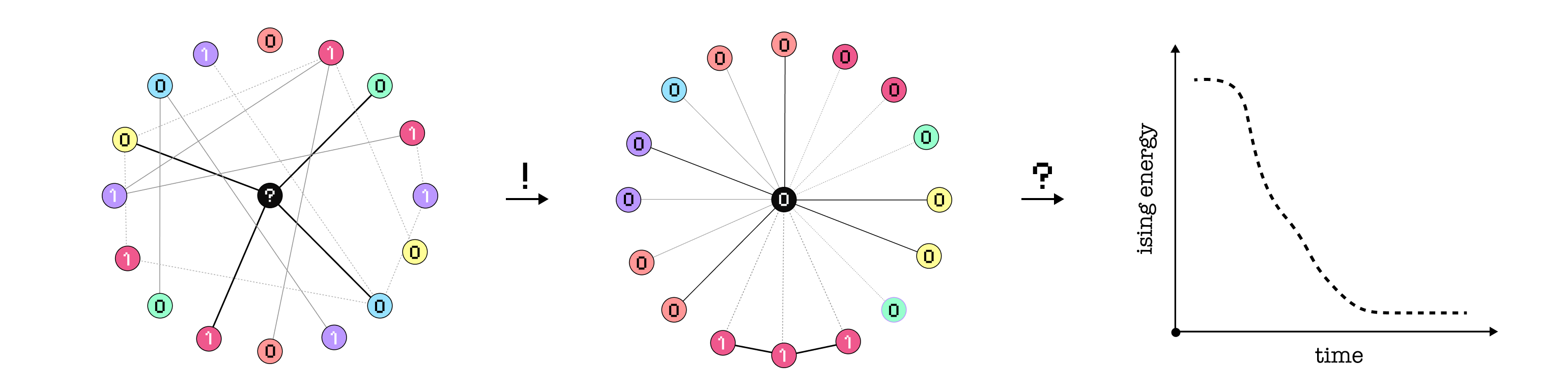}
  \caption{Neural Cellular Automata solving density classification, displaying Ising-like behaviour.}
  \Description{}
  \label{fig:teaser}
\end{teaserfigure}


\maketitle

\section{Introduction}
\label{sec:intro}
 
Many collectives reach decisions with no central controller: flocks settle on a
heading, colonies choose a nest, and distributed systems agree on a value, all
through local interactions among simple agents \cite{reynolds, vicsek}. A long line of work in collective
intelligence and multi-agent systems studies how such global agreement emerges
from local rules \cite{bonabeau}. Almost universally, these models assume that the agents share a
common communication protocol: a signal emitted by one agent means the same thing
to its neighbours.
 
Real collectives are not so uniform. The members of a group often encode and
interpret information differently, whether because of language, training,
sensory apparatus, or culture \cite{hong}. When agents communicate through mismatched
protocols, coordination must overcome not only the difficulty of the task but also
the loss incurred in translation between them. How this protocol heterogeneity
affects a collective's ability to reach agreement is studied here.
 
Our agents are the cells
of a Neural Cellular Automaton \cite{niklasson2021self-organising}, each running an identical learned local rule and
communicating only with immediate neighbours. The collective must solve density
classification, a canonical consensus task in which every cell must agree on
whether the initial configuration held more $1$s or more $0$s, a global property
no cell can see locally \cite{belew}. We partition the cells into ``languages'': sub-populations
that read a foreign neighbour's message through a fixed translation, parameterised
by a single linguistic distance that interpolates from a shared language to full
foreignness. Holding the task, the rule, and the training objective fixed and
varying only this distance (together with the size of the minority-language group
and the size of the collective) isolates the effect of communication-protocol
mismatch\footnote{We emphasise that ``language'' here is an abstract stand-in for a
difference in protocol; we make no claim to model any specific human language.}.
 
This paper makes the following contributions.
\begin{itemize}
  \item We introduce a fully controllable model of \emph{heterogeneous
  communication} in a collective: a learned local rule shared by all agents,
  with a tunable distance between the protocols of different sub-populations
  (\S\ref{sec:method}).
  \item We show that linguistic distance slows consensus, with a cost that
  accelerates with distance and, more importantly, \emph{compounds with the size
  of the collective} (\S\ref{sec:res-distance}).
  \item We show that the disruption produces mild between-group divergence rather
  than full fragmentation, and that a collective trained under diverse protocols
  is robust to mismatch that a homogeneously trained one is not
  (\S\ref{sec:res-robust}, \S\ref{sec:res-secondary}).
  \item We confirm that these findings generalise from a ring to a
  two-dimensional grid, and we give a statistical-physics reading in which
  consensus is Ising relaxation and a foreign-language region is a boundary
  defect (\S\ref{sec:res-2d}, \S\ref{sec:res-ising}).
\end{itemize} 
Taken together, the results indicate that distance between communication protocols is, on its own, sufficient to produce two qualitative patterns in a model collective: an upfront coordination cost that grows with protocol mismatch, and an advantage for collectives whose shared rule was learned under varied protocols, which continue to perform well when they meet a mismatch they never saw during training.

\section{Related Work}
\subsection{Collective Consensus and Density Classification Task} Reaching global agreement from purely local interactions is a long studied problem in distributed and multi-agent systems, where it is framed as the consensus problem \cite{LiTanconsensussurvey}. The density (or majority) classification task is its canonical instantiation in cellular automata: cells must collectively decide a global property of the initial configuration using only local updates. A foundational result is that no two-state cellular automaton solves the task perfectly for all inputs \cite{belew}, which is precisely why it remains a useful benchmark for approximate collective computation. Strong approximate rules were found first by hand \cite{GacKurLev78} and later by evolutionary search, which discovered rules that perform global computation through emergent travelling "particles" \cite{Mitchell2, mitchell3} that carry and compare local density estimates \cite{andre, MITCHELL1994361}. Recent work continues to characterise these rules, including their behaviour under noise, with explicit connections to collective intelligence in biological systems \cite{Challa2024-uzAlifenoise}. Our task is this classic problem; our contribution is a study of how heterogeneous communication among the agents affects the consensus they reach. The task has been attempted with Neural Cellular Automata under different rules and settings \cite{SIPPER1996193, capcarrere}.

\begin{figure}[t]
  \centering
  \includegraphics[trim=0cm 0cm 0cm 1.8cm, clip=true, width=0.7\linewidth]{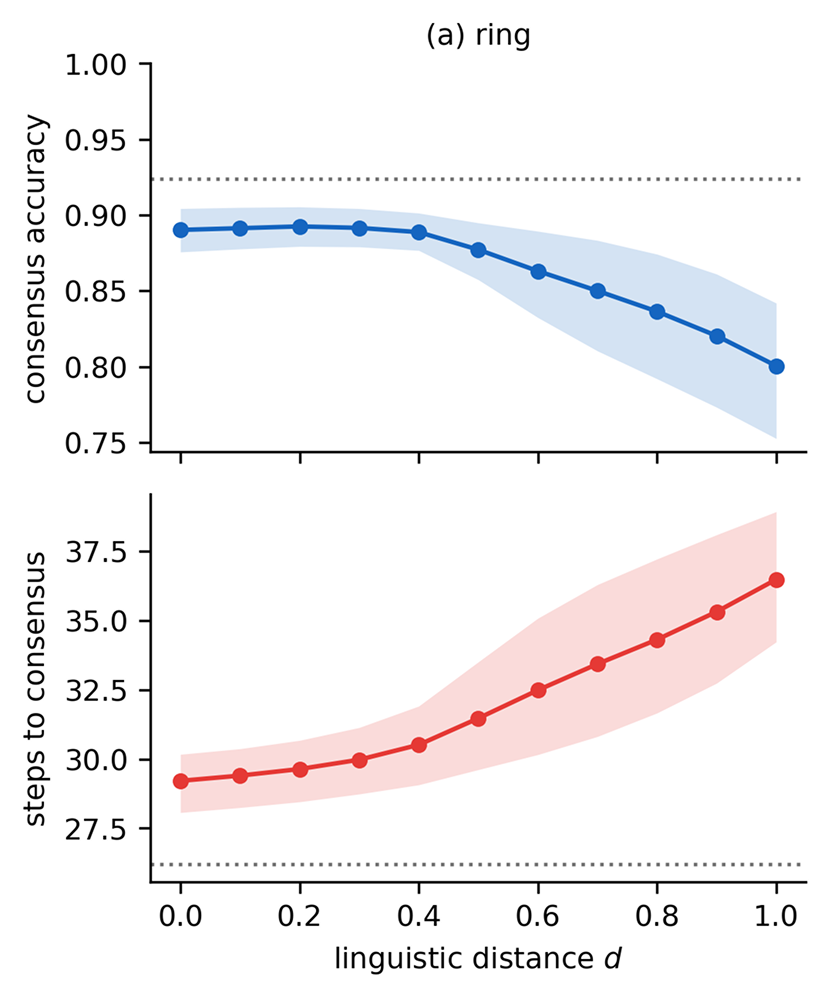}
  \caption{Consensus accuracy and time-to-consensus for a ring-topology as a function of linguistic
  distance $d$ (with minority fraction $f=0.3$, $N=50$). Time-to-consensus rises and accelerates with
  distance; accuracy declines monotonically in the high-distance regime.}
  \label{fig:1}
\end{figure}

\begin{figure}[t]
  \centering
  \includegraphics[trim=0cm 0cm 0cm 1.8cm, clip=true, width=0.7\linewidth]{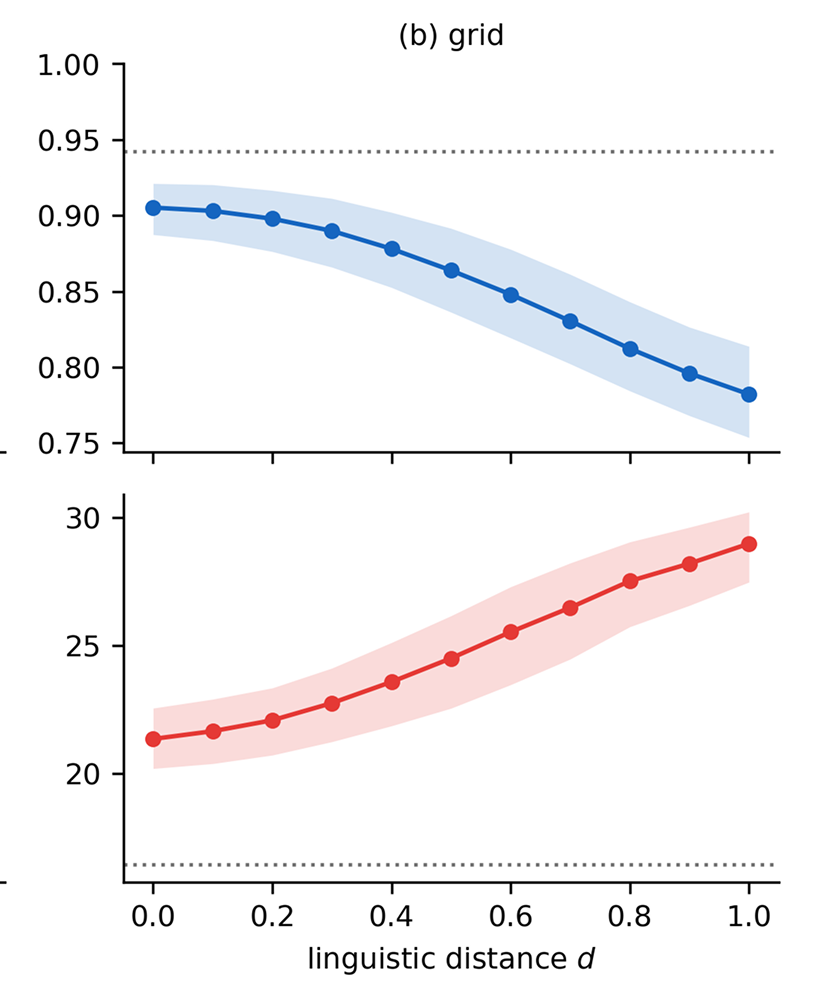}
  \caption{Consensus accuracy and time-to-consensus for a grid-topology as a function of linguistic
  distance $d$ (with minority fraction $f=0.3$, $N=50$).}
  \label{fig:2}
\end{figure}

\begin{figure*}[t]
  \centering
  \includegraphics[trim=0cm 0cm 0cm 0.0cm, clip=true, width=\linewidth]{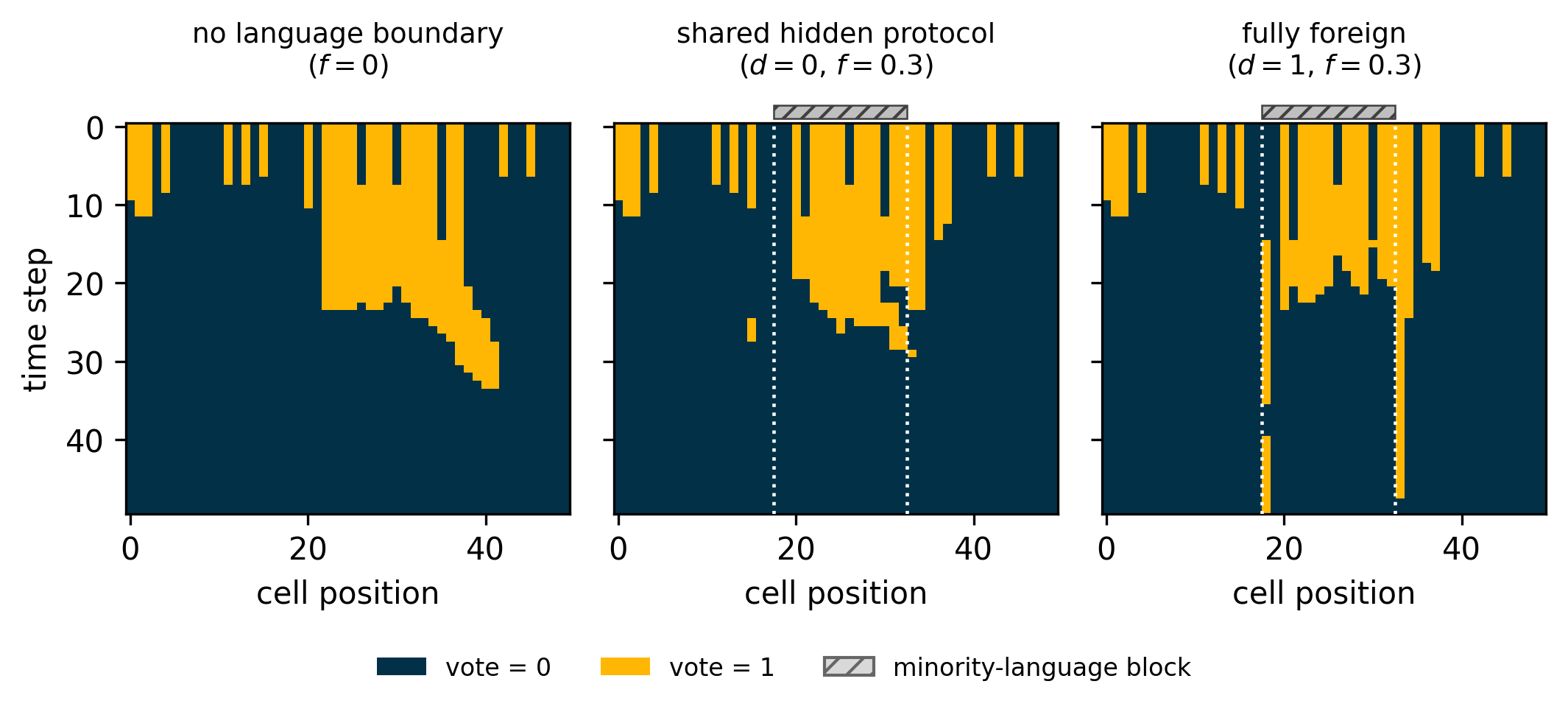}
  \caption{Space--time diagrams of a representative run under the same initial
  condition on a ring topology. Left: same language ($d=0$) resolves to a clean consensus block.
  Right: a fully foreign minority ($d=1$, dotted region) traps residual
  disagreement at the language boundary and resolves more slowly. Here, '$d$' is linguistic distance, and '$f$' is the minority fraction.}
  \label{fig:3}
\end{figure*}

\subsection{Neural Cellular Automata}
NCAs replace the fixed local rule of a classical cellular automaton with a small, trainable neural network shared across all cells \cite{mordvintsev2022growingisotropicneuralcellular}, and have demonstrated robust self-organisation across a range of tasks: growing and regenerating target morphologies \cite{mordvintsev2022growingisotropicneuralcellular}, synthesising textures \cite{sudhakaran2021growing3dartefactsfunctional, niklasson2021self-organising}, performing algorithmic and abstract reasoning \cite{kevinxu}, and executing general computation such as matrix operations \cite{bena}. The work most closely related to ours is self-classifying MNIST \cite{randazzo2020self-classifying}, in which cells reach a distributed consensus about which digit they collectively form, and subsequent analyses of the stability and adversarial robustness of that consensus \cite{Stovold_2025, randazzo2021adversarial}. These studies establish that NCAs can reach collective agreement, but they assume a homogeneous population in which every cell shares one communication protocol. NCAs have also been positioned as a unifying, computationally lean model of collective intelligence spanning biology and AI \cite{Hartl2025-be}, the framing we find useful.

\begin{figure*}[t]
  \centering
  \includegraphics[trim=0cm 0cm 0cm 0.0cm, clip=true, width=\linewidth]{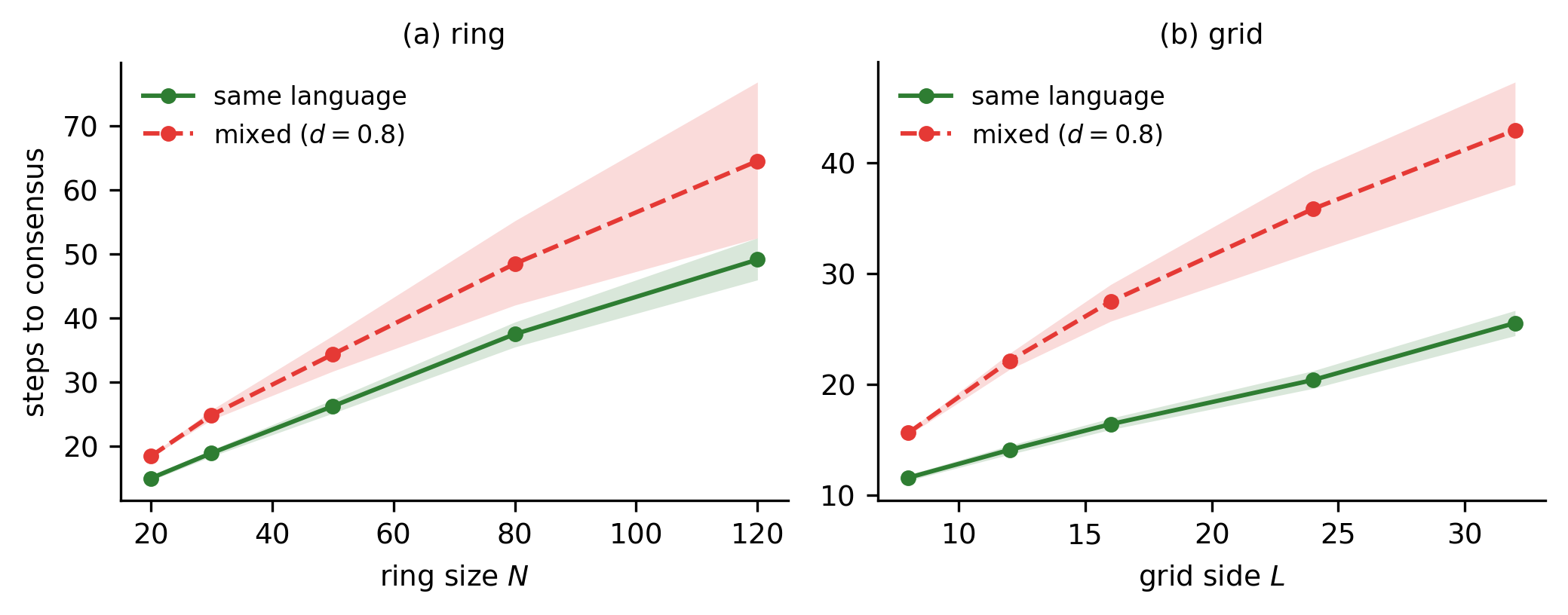}
  \caption{Time-to-consensus versus ring/grid size for the same-language and
  mixed-language ($d=0.8$, $f=0.25$) conditions. The penalty from linguistic
  heterogeneity widens with the size of the collective.}
  \label{fig:4}
\end{figure*}

\subsection{Heterogeneity and Communication in Collectives}
Models of opinion and consensus dynamics \cite{woolley, reidl} study how local interaction rules drive a population toward agreement or disagreement, including bounded-confidence models in which agents only influence one another when sufficiently similar \cite{Deffuant2000MixingBA, Hegselmann, beckerjoshua}. A complementary line in formal consensus dynamics shows that coordination slows when agents cannot infer the group's eventual convergence point from individual exchanges\cite{almaatouq, stephanie}, with the rate depending on interaction structure \cite{tsitsiklis}. Within NCAs, giving cells distinct identities has been shown to affect stability \cite{Stovold_2025}, but identity there is not framed as a communication protocol and is not varied along a distance axis. In human group research, the effect of heterogeneity is well studied but two-sided: diverse and multicultural groups often incur coordination and conflict costs in process while, under the right conditions, achieving more robust or higher-quality outcomes \cite{oetzel}. Our toy model is consistent with both sides of this picture, exhibiting an upfront coordination cost from protocol distance (§4.1, §4.3) alongside a robustness benefit from prior exposure to diversity (§4.4). We position our contribution as isolating a minimal mechanism, distance between communication protocols, that is sufficient on its own to reproduce these qualitative patterns, without invoking anything language-specific.

\section{Method}
\label{sec:method}

\subsection{Overview}
We study how heterogeneity in communication protocol affects a collective's
ability to reach agreement. Our system is a one-dimensional Neural Cellular
Automaton (NCA): a ring of cells that each run an identical, learned local
update rule and communicate only with their immediate neighbours. The collective
must solve the \emph{density classification} task, agreeing unanimously on
whether the initial configuration contained more $1$s or more $0$s, a global
property that no single cell can observe directly. We introduce ``languages'' as
sub-populations of cells that encode and read their neighbours' messages
differently, and we vary the distance between languages, the size of the
minority-language community, and the size of the collective. Throughout,
``language'' is an abstract stand-in for a difference in communication protocol;
we make no claim to model any specific human language.

\subsection{The collective and its task}
The system is a ring of $N$ cells with periodic boundaries. Each cell holds a
state vector of $C$ channels. Channel $0$ is the cell's \emph{vote}, decoded
through a sigmoid into a value in $[0,1]$; the remaining $C-1$ channels are
hidden and carry coordination signals. We use $C = 8$ throughout.

At initialisation, each cell is assigned a bit so that the ring has density
$\rho$ (the fraction of $1$s). The cell's vote channel is set to $+3$ if its bit
is $1$ and $-3$ if its bit is $0$ (so the initial decoded vote is near $0.95$ or
$0.05$), and its hidden channels are set to zero. The target for every cell is
the global majority bit, $y = \mathbb{1}[\rho > 0.5]$. To keep the majority label
unambiguous and the task non-trivial, we sample $\rho \sim \mathcal{U}(0.35,
0.65)$ and reject any configuration within $0.04$ of a perfect tie. Densities
near $0.5$ make the task genuinely hard: a cell cannot infer the global majority
from its local neighbourhood, so the collective must propagate and integrate
information to succeed.

Density classification is a natural testbed for collective consensus because it
is provably impossible to solve perfectly with a single fixed local
rule \cite{belew, capcarrere}, which reframes the question from \emph{is it solved}
to \emph{how well, and how fast, does the collective approximate agreement};
precisely the regime in which heterogeneity effects are interesting.

\begin{figure*}[t]
  \centering
  \includegraphics[trim=0cm 0cm 0cm 0.0cm, clip=true, width=\linewidth]{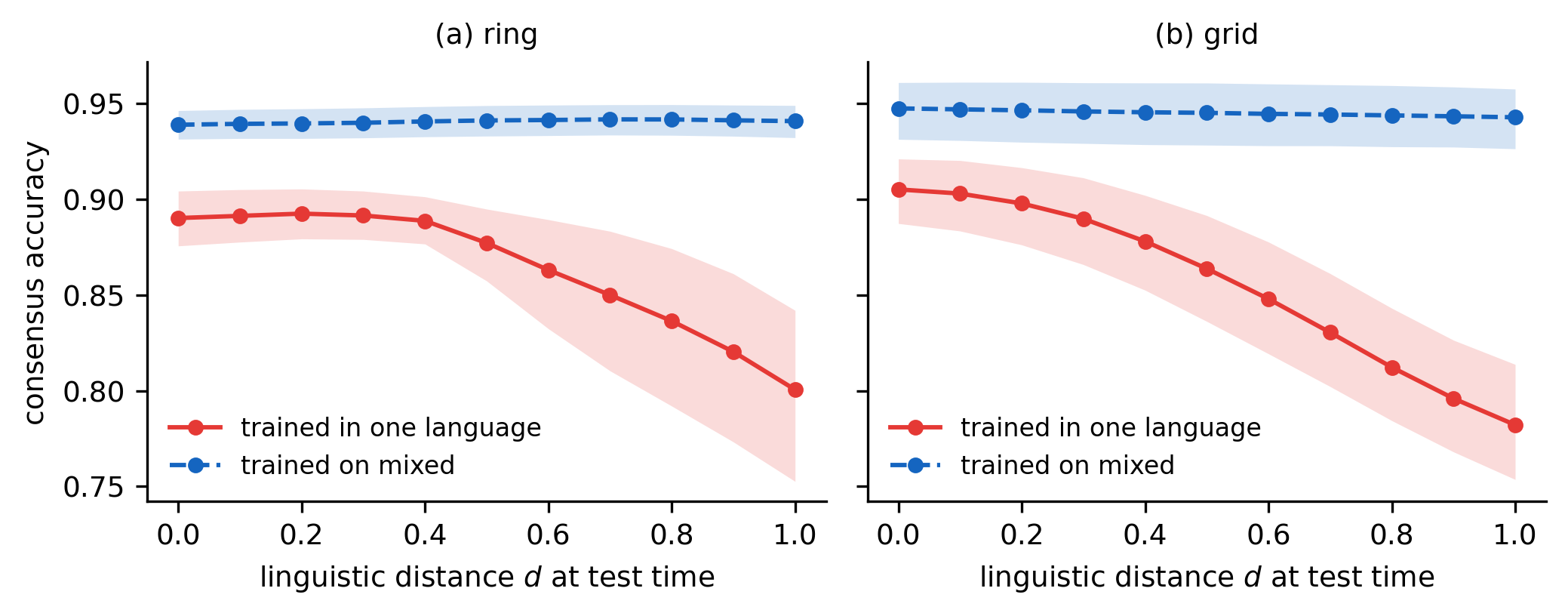}
  \caption{Consensus accuracy versus test-time linguistic distance for models
  trained in a single language and one trained on mixed compositions. Prior
  exposure to diversity confers robustness to unseen protocol mismatch.}
  \label{fig:5}
\end{figure*}

\subsection{The update rule}
All cells share a single update network $f_\theta$, the defining constraint of an
NCA. Each cell perceives its own full state together with its two neighbours'
states, and $f_\theta$ maps this perception to a state increment. With radius-$1$
neighbourhoods the perception vector has dimension $3C$: the cell's own $C$
channels, plus each neighbour's vote and $C-1$ hidden channels. The network is a
two-layer perceptron,
$\mathrm{Linear}(3C, 64) \to \mathrm{ReLU} \to \mathrm{Linear}(64, C)$,
roughly $2.2\times 10^{3}$ parameters in total. Following standard NCA practice,
the final layer is initialised to zero so the model begins as the identity
(no-op) update and grows its dynamics during training. The update is residual,
\begin{equation}
  s_i \;\leftarrow\; s_i + f_\theta(p_i),
  \label{eq:update}
\end{equation}
where $p_i$ is the perception vector of cell $i$, applied synchronously to all
cells. A cell's decision at any step is $\mathbb{1}[\,s_i^{(0)} > 0\,]$, where
$s_i^{(0)}$ denotes its vote channel.

\subsection{Languages as translation between protocols}
\label{sec:languages}
We partition the $N$ cells into language groups. In the main experiments there
are two: a majority language and a minority language occupying a contiguous block
of the ring (a geographic community); we also support a dispersed assignment as a
robustness check.

When a cell reads a neighbour of the \emph{same} language, it reads the
neighbour's channels directly. When it reads a neighbour of a \emph{different}
language, the neighbour's hidden channels pass through a translation matrix
\begin{equation}
  T(d) \;=\; (1-d)\,I + d\,M,
  \label{eq:translation}
\end{equation}
where $M$ is a fixed random permutation of the hidden channels (a foreign
``code'') and $d \in [0,1]$ is the \emph{linguistic distance}: $d=0$ recovers the
identity (the two groups share a language), while $d=1$ reads every foreign
message through the full permutation $M$. Intermediate $d$ interpolates between
mutual intelligibility and full foreignness. The same permutation $M$ is used
everywhere, so $d$ is the sole knob controlling protocol mismatch. Because $M$ is
a permutation, $T(1)$ is orthogonal: a fully foreign translation reorders which
channel carries which signal without amplifying or attenuating the message, so
any measured effect stems from miscommunication rather than from foreign messages
being louder or quieter.

Crucially, the vote channel is also subject to translation across languages: a
foreign neighbour's vote is read with its sign inverted. Without this, every cell
could read every neighbour's current vote in the clear, and the collective could
approximate the answer without relying on the hidden coordination channels,
leaving linguistic distance with almost nothing to disrupt. Scrambling the vote
across languages forces the hidden channels to carry genuine coordination, so
that protocol mismatch has a real cost.

\subsection{Metrics}
\label{sec:metrics}
For each run we record the vote trajectory over time and compute four quantities.
\emph{Accuracy} is the fraction of cells holding the correct majority vote at the
final step. \emph{Success} is the fraction of runs in which every cell agrees
\emph{and} the agreed value is correct. \emph{Time-to-consensus} is the first
step at which at least $95\%$ of cells share a vote, censored at the rollout
length if never reached. \emph{Fragmentation} is the absolute difference between
the two language groups' mean final votes, a measure of whether the groups settle
on different answers; it is zero by construction in the single-language
condition.

\subsection{Training and evaluation}
\label{sec:training}
We train two models. The \emph{monolingual} model is trained entirely in the
single-language setting ($d=0$). The \emph{multilingual} model is trained on
randomly mixed compositions: each minibatch samples a linguistic distance
$d \sim \mathcal{U}(0,1)$ and a minority fraction $f \sim \mathcal{U}(0, 0.5)$, so
the shared rule is exposed to a range of protocol mismatches during learning.
Both share the same architecture and the same fixed permutation $M$.

Each model is trained by backpropagation through time over a rollout of $N$
steps, with a binary cross-entropy loss between each cell's vote logit and the
majority bit, averaged over all cells and over the final $K=10$ steps:
\begin{equation}
  \mathcal{L}
  \;=\;
  \frac{1}{K}\sum_{t=T-K+1}^{T}\;\frac{1}{N}\sum_{i=1}^{N}
  \mathrm{BCE}\!\left(\sigma\!\big(s_i^{(0),(t)}\big),\, y\right),
  \label{eq:loss}
\end{equation}
where $\sigma$ is the sigmoid and $T$ the rollout length. We optimise with Adam
(learning rate $10^{-3}$, gradient norm clipped to $1$) for
$\approx 600$ iterations 
at batch size $\approx 32$ 
on $N=50$.

At evaluation we sweep linguistic distance, minority fraction, and ring size.
Each condition is averaged\footnote{Each point is the mean over 4,800 runs (5 independently trained models × 15 initial-condition seeds × 64 configurations per seed); shaded bands are 95\% bootstrap CIs computed hierarchically, resampling over training replicates.} over $\approx 15$ 
independent initial-condition seeds, each a batch of $\approx 64$ 
random configurations, and we report the mean and standard deviation across
seeds. Unless stated otherwise, the distance sweep fixes $f = 0.3$ and $N = 50$,
and the fraction sweep fixes $d = 0.8$ and $N = 50$.

\begin{figure*}[t]
  \centering
  \includegraphics[trim=0cm 0cm 0cm 0cm, clip=true, width=\linewidth]{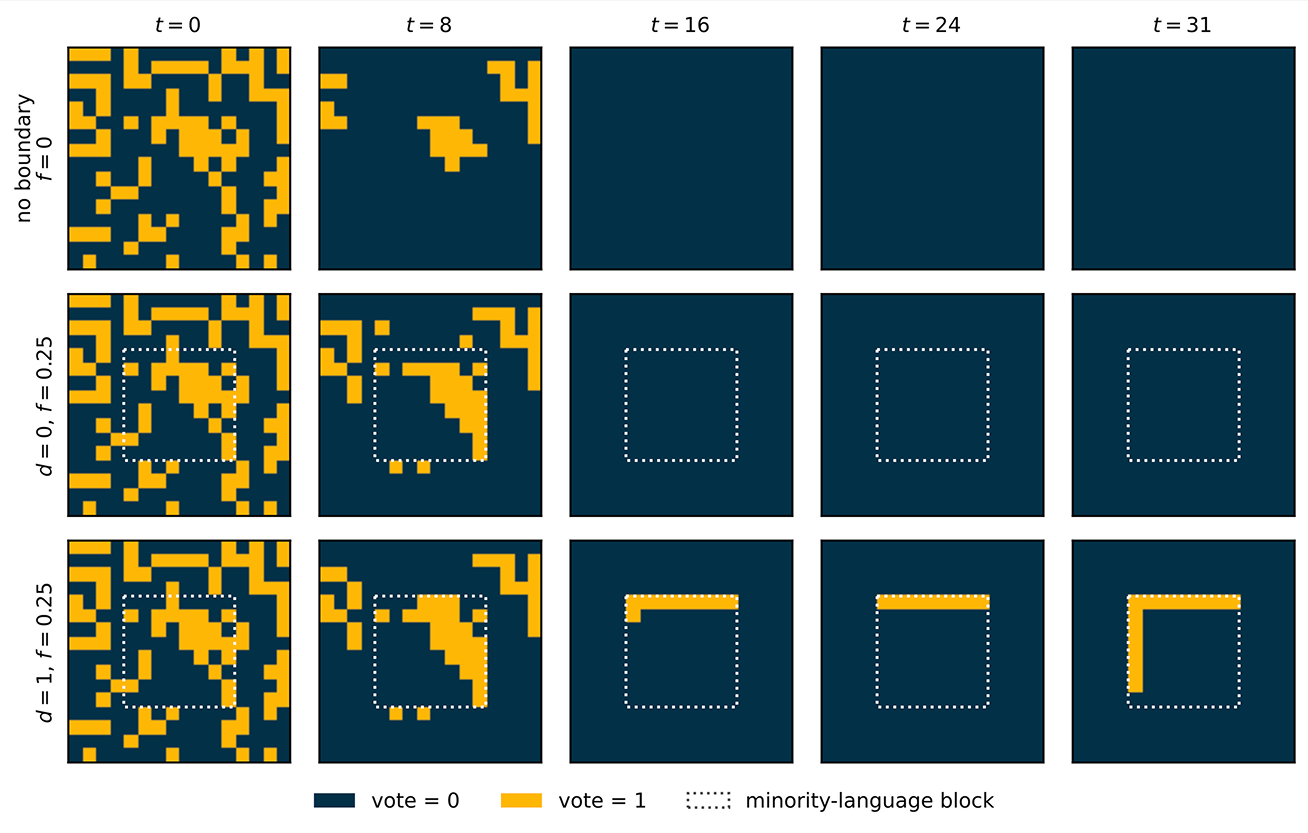}
  \caption{Grid snapshots over time (top: same language; bottom: a fully foreign
  region, dotted box) from an identical initial condition. The same-language
  collective clears to consensus, while disagreement remains trapped at the
  minority-language boundary.}
  \label{fig:6}
\end{figure*}


\section{Results}
\label{sec:results}

We organise the results around four questions: does linguistic distance impede
consensus (\S\ref{sec:res-distance}), what does the disruption look like
dynamically (\S\ref{sec:res-viz}), how does it scale with the size of the
collective, and can prior exposure to diversity confer
robustness (\S\ref{sec:res-robust})? We then report secondary observations on
minority size and fragmentation (\S\ref{sec:res-secondary}), confirm that the
findings generalise from the ring to a two-dimensional grid
(\S\ref{sec:res-2d}), and give a statistical-physics reading of the dynamics
(\S\ref{sec:res-ising}).

\subsection{Linguistic distance slows consensus, and the cost accelerates}
\label{sec:res-distance}
Figures~\ref{fig:1} and ~\ref{fig:2} show the monolingual collective's behaviour as
linguistic distance increases from $d=0$ (a single shared language) to $d=1$
(fully foreign minority), at fixed minority fraction $f=0.3$ and ring size
$N=50$. Time-to-consensus for the ring topology rises monotonically, 
and the curve steepens at larger distances rather than rising linearly. Accuracy
is comparatively robust but bends downward in the high-distance regime, 
and the monotonic decline across all sampled distances indicates a trend
rather than noise. The collective still reaches agreement, but increasingly
slowly, and slightly less reliably, as the two groups' protocols diverge.

The interpretation is that translation loss (Eq.~\ref{eq:translation}) degrades
exactly the hidden coordination signals the collective relies on to propagate
information about the global majority. With the vote channel itself rendered
language-specific, a foreign neighbour's signal is actively misleading rather
than merely uninformative, and the collective must do more work, over more steps,
to reach the same agreement.

\begin{figure*}[t]
  \centering
  \includegraphics[trim=0cm 0cm 0cm 0cm, clip=true, width=\linewidth]{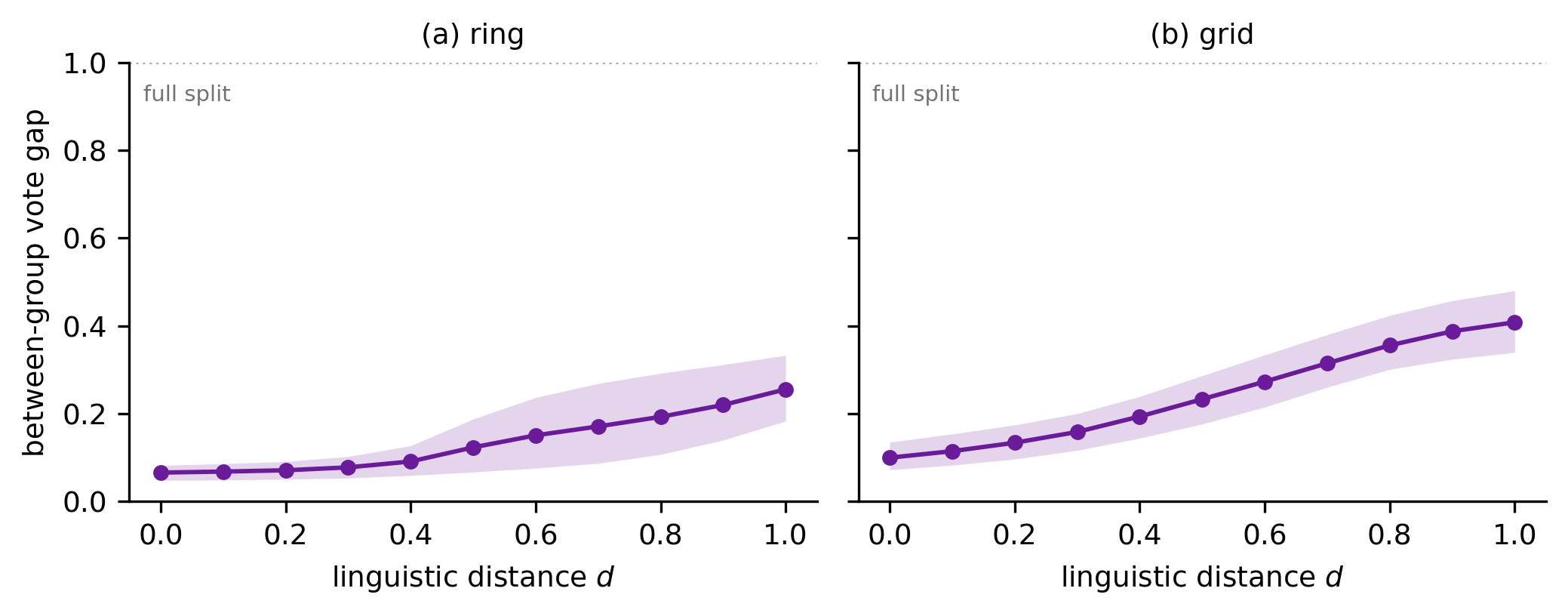}
  \caption{Plots for between-group vote gaps for both topologies. Y-axis = |mean(group $A$ votes) - mean(group $B$ votes)|. The consensus accuracy is not in question here, but the fraction of minority votes.}
  \label{fig:frag}
\end{figure*}

\begin{figure*}[t]
  \centering
  \includegraphics[trim=0cm 0cm 0cm 1.4cm, clip=true, width=\linewidth]{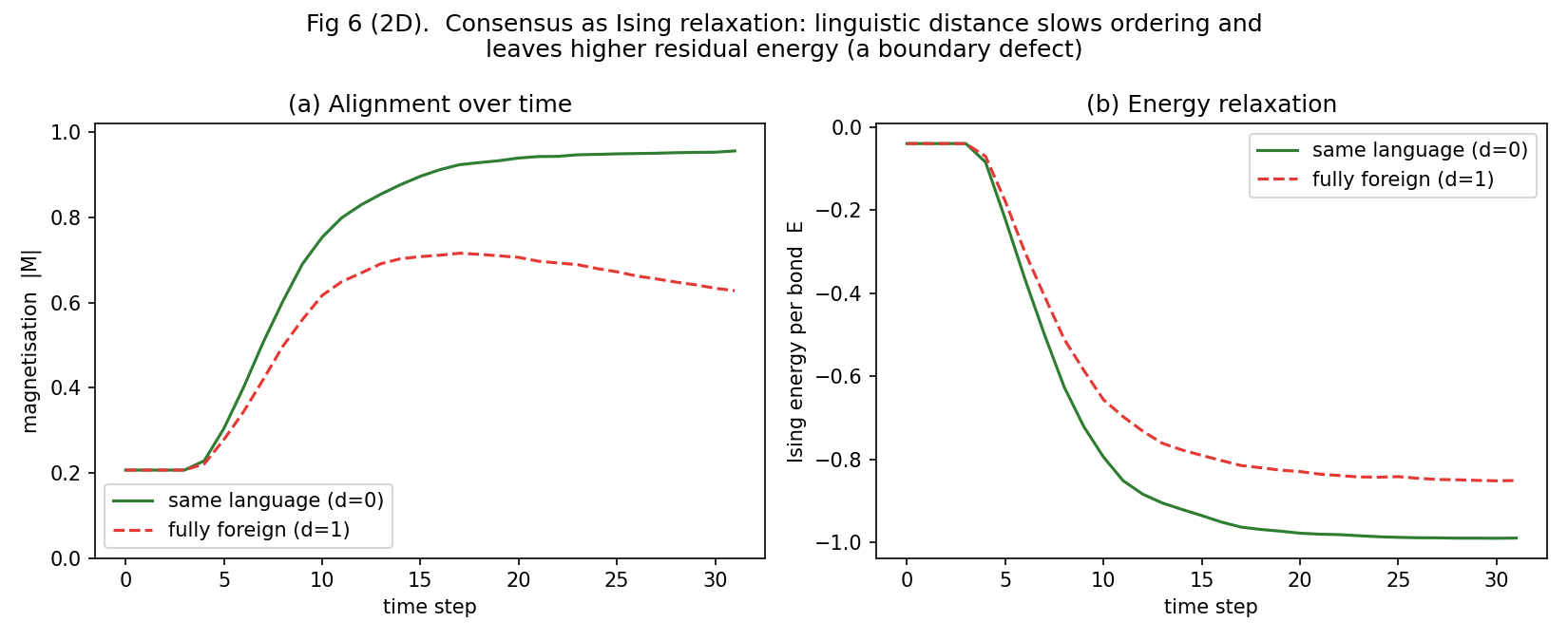}
  \caption{Consensus as Ising relaxation. (a)~Magnetisation $|M|$ over time:
  the same-language system aligns fully while the fully-foreign system plateaus.
  (b)~Ising energy per bond over time: the same-language system relaxes toward
  the ground state, while the fully-foreign system settles at higher residual
  energy, the signature of a boundary defect.}
  \label{fig:ising}
\end{figure*}

\subsection{Visualising the disruption}
\label{sec:res-viz}
Figure~\ref{fig:3} makes the mechanism visible for ring-topology (while Figure~\ref{fig:6} also does, but for a grid topology, which will be discussed in §~\ref{sec:res-2d}). Each panel is a
space--time diagram of a single representative run from an identical initial
condition: cell position on the horizontal axis, time descending on the vertical
axis, and colour encoding each cell's vote. In the same-language panel ($d=0$),
local regions of agreement form and the majority value sweeps across the ring,
resolving into a uniform consensus block. In the fully-foreign panel ($d=1$), the
same initial configuration resolves more slowly, and residual disagreement
concentrates around the minority-language region, where translation loss
repeatedly disrupts the propagation of coordination signals. The boundary between
language communities acts as a partial barrier \cite{Barkoczi2016}to the flow of information, the
spatial signature of the slowdown quantified in Figure~\ref{fig:1}.

\subsection{The cost compounds with the size of the collective}
\label{sec:res-scalin}
Figure~\ref{fig:4} compares time-to-consensus for the single-language and
mixed-language conditions across ring sizes $N \in \{20, 30, 50, 80, 120\}$. Both
grow with $N$, as expected: information must traverse a larger ring to reach
global agreement. The key observation is that the gap between the two conditions
widens with size. Linguistic heterogeneity therefore does not impose a fixed overhead; its
cost compounds as the collective grows. This is the most consequential of our
findings for real systems, where coordination problems are typically larger than
any toy ring: protocol mismatch that is tolerable in a small group can become a
meaningful drag at scale.

\subsection{Experiencing diversity confers robustness}
\label{sec:res-robust}
Figure~\ref{fig:5} contrasts the two trained models under increasing
test-time linguistic distance. The monolingual model, which never encountered
protocol mismatch during training, degrades as distance grows, with accuracy
declining. 
The multilingual model, trained on randomly mixed compositions, is both more
accurate overall 
and essentially flat across the entire range of test-time distances. A collective
whose shared rule has been shaped by exposure to diverse protocols is robust to
protocol mismatch it has not specifically seen, whereas one shaped in a
homogeneous setting is not. The absolute gap is modest, consistent with the small
effect sizes throughout; the qualitative separation, however, is clear and
monotonic.

\subsection{Minority size and fragmentation}
\label{sec:res-secondary}
We find that the \emph{distance} of the minority language matters more than the
\emph{size} of the minority community. Varying the minority fraction $f$ at fixed
distance $d=0.8$ produces no clear monotonic trend in either accuracy or
time-to-consensus within the noise of our seeds; the presence of a sufficiently
foreign sub-population, rather than its exact size, drives the slowdown.
Relatedly, Figure~\ref{fig:frag} shows that between-group fragmentation rises with
linguistic distance, but remains small in absolute terms across the whole range. The two language
groups diverge mildly in their settled votes; they do not split into separate,
internally-agreeing camps. At these parameters the collective bends under
protocol mismatch without fracturing.

\begin{figure*}[t]
  \centering
  \includegraphics[trim=0cm 0cm 0cm 1.0cm, clip=true, width=\linewidth]{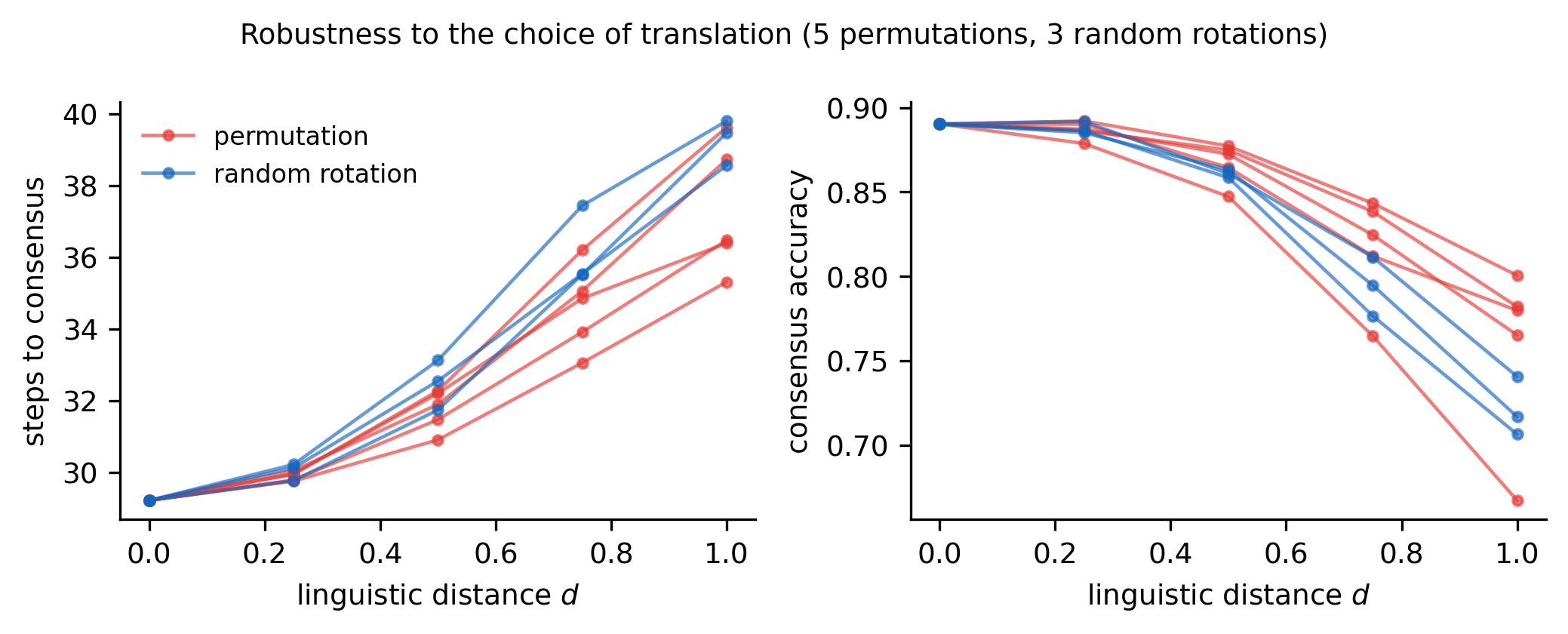}
  \caption{Plots for the accuracy and consensus over different permutations and random shuffles. There are 5 different permutations, and 3 different random runs.}
  \label{fig:10}
\end{figure*}

\subsection{Generalisation to two dimensions}
\label{sec:res-2d}
To test whether these effects are an artefact of the ring topology \cite{grattarola}, we repeat the
experiments on an $L \times L$ grid with a von Neumann (four-neighbour)
neighbourhood, the two-dimensional analogue of the ring's radius-$1$
connectivity. The same task, update rule, and language mechanism are used; only
the connectivity changes. All four findings carry over: linguistic distance slows
consensus and reduces accuracy, the penalty again compounds with system size, the
multilingual model remains robust where the monolingual one degrades, and
fragmentation stays modest. Figure~\ref{fig:6} shows grid snapshots over
time for the same initial condition: the same-language grid clears to a uniform
consensus, while a fully foreign region traps a stubborn pocket of disagreement
exactly at the minority-language block. The agreement of the ring and grid
results indicates that the phenomenon depends on the presence of a communication
boundary, not on any specific topology.

\subsection{A statistical-physics view: consensus as Ising relaxation}
\label{sec:res-ising}
The grid setting admits a natural reading in terms of the Ising model. Mapping
each vote to a spin $\sigma_i = 2\hat{y}_i - 1 \in \{-1, +1\}$, the collective's
alignment is the magnetisation $M = \tfrac{1}{N}\sum_i \sigma_i$, and the quality
of local coordination is the energy per bond $E = -\langle \sigma_i \sigma_j
\rangle$ over nearest neighbours. Reaching consensus corresponds to the lattice
ordering: $|M| \to 1$ and $E$ relaxing toward its ground state. Under this view, a
foreign-language region behaves like a defect that frustrates the bonds crossing
it.

Figure~\ref{fig:ising} plots both quantities over time for the same-language and
fully-foreign conditions. In the same-language case the magnetisation climbs
toward and the energy relaxes to near the ordered ground state. 
In the fully-foreign case the system orders only partially, with magnetisation
plateauing 
and the energy settling at a higher residual value, 
the signature of frustrated bonds along the language boundary. The same
phenomenon that appears as slower consensus in
\S\ref{sec:res-distance}--\S\ref{sec:res-scalin} thus appears here as incomplete
relaxation to a higher-energy state. We present this correspondence as an
interpretive lens rather than a formal equivalence: it connects our learned
collective to a well-understood model of ordering dynamics and clarifies why a
communication boundary impedes global agreement.

\section{Discussion}
\label{sec:discussion}
 
Our central observation is that a single, interpretable quantity, the distance
between two sub-populations' communication protocols over different permutations and random shuffles (as displayed in Figure~\ref{fig:10}), is enough to impose a
measurable and structured cost on collective consensus. The cost is not a fixed
overhead: it accelerates as protocols diverge and compounds as the collective
grows, so a mismatch that is harmless in a small group can become a meaningful
drag at scale. At the same time, the collective is more resilient than a naive
reading would suggest. It bends rather than fractures, the two groups diverging
only mildly, and prior exposure to diverse protocols during training confers
robustness to mismatch never seen before.
 
These patterns are qualitatively consistent with a substantial body of human group
research, though we draw the connection at the level of mechanism rather than
claiming to reproduce any specific human study through NCAs. Work on multicultural and diverse teams
reports a similar two-sided picture: heterogeneous groups often incur coordination
and process costs while, under the right conditions, achieving more robust or
higher-quality outcomes \cite{oetzel}. Formal models of
consensus and opinion dynamics likewise find that agreement slows when agents
cannot infer the group's eventual convergence point from individual
exchanges \cite{Hegselmann}, which is precisely what
translation loss induces here. Unlike communication accommodation \cite{giles}, where protocols adapt toward one another, our translation is fixed and no agent can reduce its own mismatch, our measured costs are a lower bound on what an accommodating collective could achieve. 
 
The Ising reading of our grid results (\S\ref{sec:res-ising}) connects the learned
collective to a well-understood model of ordering dynamics. Identifying consensus
with magnetisation and local coordination with energy relaxation makes the role of
a foreign-language region transparent: it behaves as a boundary defect that
frustrates the bonds crossing it, leaving the system in a higher-energy, partially
ordered state. We offer this as an interpretive lens rather than a formal
equivalence; making the correspondence precise, for instance by relating linguistic
distance to a defect coupling strength, is an appealing direction for future work.

\newpage
\newpage

\bibliographystyle{ACM-Reference-Format}
\bibliography{references}

\end{document}